\documentclass[lettersize,journal]{IEEEtran}
\usepackage{amsmath,amsfonts}
\usepackage{algorithmic}
\usepackage{algorithm}
\usepackage{array}
\usepackage{tabularx}
\usepackage[caption=false,font=normalsize,labelfont=sf,textfont=sf]{subfig}
\usepackage{textcomp}
\usepackage[table,xcdraw]{xcolor}
\usepackage{stfloats}
\usepackage{url}
\usepackage{verbatim}
\usepackage{tikz}
\usetikzlibrary{positioning, shapes}
\usetikzlibrary{positioning, shapes.geometric}
\usepackage{graphicx}
\usepackage{cite}
\usepackage{amsmath}
\usepackage{booktabs}
\usepackage{amssymb}
\usepackage{nomencl}
\makenomenclature
\begin{document}

\title{A Bézier Curve–Based Approach to the Convexification of the AC Optimal Power Flow Problem}

\author{Carlos Arturo Saldarriaga-Cort\'{e}s, Carlos Adrián Correa-Flórez, Maximiliano Bueno-L\'{o}pez, \textit{Senior Member, IEEE}, Maria Victoria Gasca-Segura

\thanks{C.A. Saldarriaga Cortés is a researcher in power systems at Universidad Tecnológica de Pereira, Pereira, Colombia, e-mail: casaldarriaga@utp.edu.co}
\thanks{C.A. Correa-Flórez is with the program of Electrical Engineering, Universidad Distrital Francisco José de Caldas, Bogotá, Colombia, e-mail: ccorreaf@udistrital.edu.co}
\thanks{M. Bueno-L\'{o}pez is with the program of Electrical Technology, Universidad Tecnológica de Pereira, Pereira, Colombia, e-mail: m.bueno3@utp.edu.co.}
\thanks{M.V. Gasca, G2Elab, Université Grenoble Alpes, Grenoble, France, email:Victoria.gasca@grenoble-inp.fr}

\thanks{Manuscript received December 16, 2025; revised XXXX, 2025.}}



\maketitle

\begin{abstract}
The Alternating Current Optimal Power Flow (ACOPF) problem remains one of the most fundamental yet computationally challenging tasks in power systems operation and planning due to its nonconvex, nonlinear, and multimodal nature. This paper proposes a convex reformulation of the AC power flow problem by introducing auxiliary variables to isolate nonlinear terms, applying logarithmic transformations to exploit product-sum properties, and approximating with Bézier curves using a novel convexifying butterfly-shaped function. This model is intended for assessing and operating weak power systems that face challenges with reactive power supply and overall network robustness. Its formulation closely mirrors the AC formulation, particularly regarding active and reactive power dispatch and network voltage levels. 
 
The proposed model achieves convergence on large test systems (e.g., IEEE 118-bus) in seconds and is validated against exact AC solutions. 
This convex formulation stands out not only for its mathematical transparency and intuitive structure but also for its ease of validation and implementation, making it an accessible and reliable tool for researchers and system operators for energy planning.

The numerical analysis conducted on the IEEE 118-bus system yielded average percentage errors in the state variables—specifically, the magnitudes and angles of nodal voltages—of just 0.0008\% and 0.014°, respectively, when compared with the precise AC formulation. These results underscore the high accuracy and reliability of the proposed methodology.

\end{abstract}

\begin{IEEEkeywords}
Alternating Current Optimal Power Flow (ACOPF), Bézier curve, Convex optimization.
\end{IEEEkeywords}
\printnomenclature

\section{Introduction}

The AC Optimal Power Flow (ACOPF) problem is a cornerstone of modern power network operation and planning and has become one of the most widely studied and essential tools for scheduling generation resources in power systems. It determines the optimal dispatch of active and reactive power to meet demand at minimum cost while respecting network constraints (generator limits, line capacities, voltage profiles, etc.) \cite{wood2013power}. ACOPF underpins critical functions such as market clearing by Independent System Operators (ISOs) and security-constrained dispatch, directly impacting system efficiency and reliability. Unlike simplified DC OPF approximations, ACOPF co-optimizes active and reactive power while enforcing voltage and feasibility constraints to yield physically achievable solutions. This capability is essential across various power system tasks: in operation, it enables the coordinated dispatch of generation and reactive resources to ensure feasible, secure, and economically optimal operating conditions; in planning, it facilitates the strategic placement of reactive compensation to maintain future grid reliability and voltage stability; and in market analysis, it serves as a foundation for assessing the role and potential remuneration of reactive power as an ancillary service. Therefore, accurate and convex AC formulations are indispensable for realistic, cost-effective decision-making in modern power systems \cite{yang2018opf}, \cite{huneault1991survey}.

Solving ACOPF using simple and scalable models to achieve global optimality is still an open research question, due to its nonlinear, non-convex, and multimodal nature. The full ACOPF is NP-hard~\cite{Mantovani, ACOPFisNPHARD}, with power flow equations giving rise to multiple local minima in the objective function~\cite{bienstock2019np, bukhsh2013local}. Consequently, conventional nonlinear solvers can become trapped in suboptimal solutions. Techniques based on semidefinite programming (SDP), second-order cone programming (SOCP), and quadratic convex (QC) modeling recast the problem to eliminate or bound the non-convexities, aiming to obtain the global optimum (or a tight bound on it) in polynomial time. Under certain conditions (e.g., in many radial or well-behaved networks) these relaxations produce zero duality gap, finding globally optimal solutions to the original ACOPF. Yet, ensuring convergence to a true global optimum in general meshed ill-conditioned transmission networks, such as the case for certain zones in Latinamerican systems, remains an open challenge. Mesh topologies introduce constraints (e.g., Kirchhoff’s voltage laws) that are not easily convexified, often causing relaxations to be inexact~\cite{lavaei2012zero, kocuk2016inexactness}.

Recent trends include exploring alternative problem formulations and hybrid algorithms (e.g., network flow-based models, decomposition and parallelization strategies) to handle large, complex grids. There is also a surge of interest in integrating modern computational tools: for instance, machine learning techniques have been applied to accelerate or warm-start ACOPF solvers, predict near-optimal dispatch decisions, and assist in global solution finding~\cite{cengil2022learning}. Such data-driven approaches promise orders-of-magnitude speed-ups, though challenges remain in guaranteeing feasibility and optimality. Meanwhile, the ongoing transition to high shares of renewable generation has spurred extensions of ACOPF to incorporate uncertainty and variability. Stochastic and robust OPF formulations are being developed to accommodate wind/solar forecast errors and other uncertainties~\cite{babiker2025opf}. All of the above considering the need for simplicity, in such a way that ISOs and practitioners can easily implement.

Recent studies propose various convex relaxations and transformations of the power flow equations to bound or approximate the ACOPF solution, effectively ``searching'' for global optima within a convexified problem space~\cite{low2014convex1}. Broadly, these approaches trade off optimality guarantees, computational scalability, and solution accuracy. Key challenges include handling meshed network topologies, ensuring feasibility in the original AC equations, and achieving practical solve times for large-scale systems.

One major avenue of research is to construct relaxations of ACOPF that are convex yet as tight as possible. SDP relaxations lift the quadratic power flow constraints into higher-dimensional matrices, yielding a convex SDP that provides a global lower bound on the objective~\cite{lavaei2012zero}. If the SDP relaxation is exact (i.e., the solution matrix has rank one), it recovers the true global optimum of the nonconvex OPF~\cite{low2014convex2}. However, in general meshed networks this exactness is not guaranteed. Researchers have introduced techniques to strengthen SDP relaxations, e.g., bound tightening and convex envelopes~\cite{molzahn2019qc}. 

A more tractable class is based on SOCP. In radial networks, SOCP relaxations can be exact under certain conditions~\cite{farivar2013branch}. However, in meshed transmission grids, the SOCP relaxation is typically inexact, leaving an optimality gap. Some researchers use iterative refinement techniques such as penalty convex-concave procedures~\cite{tian2017dc}, while others develop extended convex OPF formulations to improve SOCP~\cite{yuan2018soc}. These can reduce the gap and recover feasible solutions but require additional tuning and post-processing.

Similarly the \textit{QC relaxation} convexifies trigonometric and voltage-product terms using McCormick envelopes~\cite{coffrin2016qc}. It lies between SOCP and SDP in tightness and complexity, offering improved solution quality with moderate overhead. Several works have enhanced QC relaxations by strengthening envelopes and reducing auxiliary constraints~\cite{narimani2024enhancedqc, aldik2023lineqc}. These approaches are highly competitive, although they still lack global optimality guarantees and may introduce approximation error.

Other approaches blend relaxations with deterministic search. Branch-and-bound or outer approximation techniques~\cite{aigner2020milp, jabr2025oa} can solve convexified OPF subproblems at each node and converge globally. These methods support discrete decisions (e.g., topology switching), but scale poorly unless paired with efficient cut generation or warm starts. 

Some recent research proposes alternative reformulations of the AC power flow equations. For example, semi-Lorentzian mappings~\cite{gholami2022semi} aim to directly embed nonlinearities into a convex structure. These transformations show tighter bounds and improved feasibility recovery but often involve custom solvers.

Convexification techniques are being extended to settings with uncertainty, discrete controls, or multi-objective planning. Chance-constrained OPF with SDP relaxations~\cite{venzke2018chance} allows incorporating wind and solar variability. Security-constrained OPF formulations with convex feasible sets have been used to ensure $N$-$1$ compliance~\cite{farid2022security}.

\subsection{Contributions of this work} 
Despite notable advances, existing OPF models that guarantee convergence face substantial challenges in meshed networks. These models often involve detailed parameter tuning, lack scalability, and their implementation is complex.

In contrast to previous research, this work introduces a fundamentally different approach in the conceptual framework. The proposed convex formulation is based on Bézier curve transformations~\cite{LAURENT2001609} and concave penalization that ensures convexity by design. It begins by introducing auxiliary variables that isolate the nonlinear terms of the AC equations, followed by a logarithmic transformation that turns products into sums. These terms are then approximated using Bézier curves, and the resulting expressions are convexified through a \textit{Butterfly-shaped Function (BF)}—a new element introduced in this paper that enables intuitive and tractable reformulation of the solution space. The result ensures global optimality with strong fidelity to the original physics, as well as ease of understanding and implementation. Unlike QC relaxations, which rely on piecewise approximations, or convex concave procedures that require multiple iterations and risk local minima, the proposed formulation yields a transparent convex structure. The model is applied to the IEEE 118-bus system, which consists of 6,433 continuous variables and 54 binary variables. It is solved in 8.11 seconds, with negligible deviation from full AC feasibility. The model is easy to understand and implement, scalable, and suitable for supporting technical, economic, and policy decisions in various scenarios.

The main contributions of this paper can be synthesized as follows: i) introduces a novel optimal power flow model with a convex formulation that significantly reduces mathematical complexity while ensuring convergence and feasibility across large-scale power systems; ii) a mathematical and computational validation confirms that the proposed approach closely approximates exact formulation of AC power flow, demonstrating both precision and robustness. And iii) its computational efficiency and ease of implementation make it well-suited for diverse network topologies and scales. Moreover, this model proves particularly advantageous for addressing key energy challenges, such as optimizing storage capacity, planning reactive power compensation — including synchronous condensers and FACTS — and integrating high shares of renewable energy sources, among other applications.

\section{ACOPF Mathematical formulation}

The ACOPF model is defined by the set of equations \eqref{eq:FO} to \eqref{eq:thetaequal}. The objective function presented in \eqref{eq:FO} can be tailored to suit various applications, but it typically comprises the following key components:
\begin{enumerate}
\item The value of produced energy, which can be modeled linearly when representing energy market revenues (e.g., for market analysis). Alternatively, it can be represented as a quadratic cost function to capture generation costs. In the latter case, the quadratic function can be approximated with high accuracy using piecewise linearization techniques.
\item A penalty for non-attended demand (NAD).
\item A penalty term to promote voltage adherence to operator-set references, thereby enhancing voltage stability. This is typically modeled in a linear fashion, using absolute deviations enforced through inequality constraints.
\end{enumerate}

The set of equations \eqref{eq:Pbalance} to \eqref{eq:thetaequal} models the physical and operational constraints of the power system. Specifically:
\begin{itemize}
\item Equations \eqref{eq:Pbalance} and \eqref{eq:Qbalance} enforce the nodal active and reactive power balance, respectively.
\item Equations in \eqref{eq:limits} define power flows and operational constraints as functions of nodal voltages, voltage angles, and network parameters.
\item The constraints in \eqref{eq:thetaequal} represent operational limits on system variables, such as voltage magnitudes, power generation bounds, and angular differences.
\end{itemize}
\begin{gather}
\begin{split}
    \label{eq:FO}\text{FO}&=\underbrace{\Delta T \cdot \sum_k{Cg_k\cdot Pg_k}}_{\text{value of produced energy }}+\underbrace{\Delta T\cdot \sum_n{CR_n\cdot R_n}}_{\text{Penalty for non attended demand}}+\\
    &\underbrace{\sum_{n\in N^{pilot}}{CV_n\cdot \left| V_n^{ob}-v_n \right|}}_{\text{Penalty for voltage deviations}}\end{split}\\
    \begin{split}
         \label{eq:Pbalance} \textrm{s.t.} \qquad  &\sum_{k \in \Omega^{G}_{n}}{Pg_k}+\sum_{ij \in \Omega^{in}_{n}}{P_{ij}}-\sum_{ij \in \Omega^{out}_{n}}{P_{ji}} = \\&Pd_n \cdot (1-R_n)      
    \end{split}\\ 
    \begin{split}
        \label{eq:Qbalance} \sum_{k \in \Omega^{G}_{n}}{Qg_k}+\sum_{ij \in \Omega^{in}_{n}}{Q_{ij}}-\sum_{ij \in \Omega^{out}_{n}}{Q_{ji}} =\\ Qd_n \cdot (1-R_n)+\sum_{c\in \Omega^{Comp}_{n}}{Q_c^{comp}}    
        \end{split}  
    \end{gather}

\begin{gather}
\begin{split}
P_{ij}&= g_{ij}\cdot v_i^{2}-g_{ij}\cdot v_{i}\cdot v_{j}\cdot cos(\theta_{ij})
-b_{ij}\cdot v_{i}\cdot v_{j}\cdot sin(\theta_{ij})\\
P_{ji}&= g_{ij}\cdot v_j^{2}-g_{ij}\cdot v_{i}\cdot v_{j}\cdot cos(\theta_{ij})+b_{ij}\cdot v_{i}\cdot v_{j}\cdot sin(\theta_{ij})\\
 Q_{ij}&= -(b_{ij}+b_{ij}^{sh/2})\cdot v_i^{2}+b_{ij}\cdot v_{i}\cdot v_{j}\cdot cos(\theta_{ij})\\
 &-g_{ij}\cdot v_{i}\cdot v_{j}\cdot sin(\theta_{ij})\\
Q_{ji}&= -(b_{ij}+b_{ij}^{sh/2})\cdot v_j^{2}+b_{ij}\cdot v_{i}\cdot v_{j}\cdot cos(\theta_{ij})\\
 &+g_{ij}\cdot v_{i}\cdot v_{j}\cdot sin(\theta_{ij})
\label{eq:limits}
\end{split}
\end{gather}

\begin{gather}
\begin{split}
-\overline{P}_{ij} \leq P_{ij},P_{ji}\leq\overline{P}_{ij}\\
 0\leq Pg_{k}\leq\overline{Pg}_{k}\\
-\underline{Qg_{k}} \leq Qg_{k}\leq\overline{Qg}_{k}\\
\underline{V} \leq v_{n}\leq \overline{V}\\
0  \leq R_n\leq 1\\
-\overline{\theta}  \leq \theta_{ij}=\theta_i-\theta_j\leq \overline\theta
\label{eq:thetaequal}
\end{split}
\end{gather}

Let $Pg_k$, $Qg_k$ , $P_{ij}$, $Q_{ij}$, $v_{n}, \theta_{n}$, $Q_{c}^{comp}$ and $R_n$:  denote the decision variables representing generation dispatch, line flows, voltage magnitudes and angles, compensation injections, and demand not served (DNS) at each node, respectively.

The constraints outlined in \eqref{eq:limits} introduce significant nonlinearities into the model, primarily through bilinear interactions between voltage magnitudes and trigonometric functions of angular differences. These characteristics are intrinsic to the AC power flow equations, contributing to the non-convexity of the ACOPF problem. Consequently, the solution space exhibits multiple local optima, and conventional nonlinear programming techniques often converge to these local optima, providing no assurance of achieving global optimality.

Furthermore, nonconvexity hinders both global optimality and scalability, motivating the use of convex reformulations. Existing approaches often face challenges such as relaxation gaps, high computational complexity, or poor accuracy in capturing real power flow behavior.

In this context, the following section introduces a novel mathematical approach that effectively tackles the challenges associated with the ACOPF problem. This method utilizes Bézier curve representations alongside the BF to reformulate the non-convex aspects of ACOPF into a fully convex optimization model characterized by quadratic terms. This transformation guarantees global convergence while maintaining high fidelity to the AC power flow equations. The formulation demonstrates robust performance across various network topologies and voltage levels, including meshed high-voltage transmission systems, consistently achieving high accuracy. Moreover, it offers a scalable and computationally efficient alternative to traditional methods, all while upholding the physical fidelity of the model.

\section{Convex reformulation of the ACOPF: The BF-ACOPF}

To address the nonlinearities in the equations from (4), the following methodology is proposed. First, the nonlinearities of the original optimal power flow model are isolated through the following variable changes: 
\begin{gather}
    \label{eq:seis} \varphi_i = v_i^{2}\\ 
\label{eq:siete} w_{ij}=v_i\cdot v_j \\ 
\label{eq:ocho} \alpha_{ij} = w_{ij}\cdot cos(\theta_{ij})\\ 
\label{eq:nueve} \gamma_{ij} = w_{ij}\cdot sin(\theta_{ij}) 
\end{gather}

A logarithmic transformation is subsequently applied to equations \eqref{eq:seis} - \eqref{eq:ocho}, which are constrained to reach non-zero and positive values. To further streamline the formulation, a second variable transformation is introduced, effectively reducing the nonlinearities and enabling a simpler relationship between pairs of variables. This process yields the following resulting equations:
\begin{gather}
\label{eq:Lphi} L\varphi = 2\cdot Lv_{i}\\
\label{eq:Lomega} Lw_{ij}=Lv_i+ Lv_j \\
\label{eq:Lalpha} L\alpha_{ij} = Lw_{ij}\cdot LCt_{ij}\\
\label{eq:Lvi}Lv_{i} = ln (v_i)\\
L\varphi_{i} = ln (\varphi_i)\\
Lw_{ij} = ln (w_{ij})\\
\label{eq:Lalphai} L\alpha_{ij} =ln (\alpha_{ij})\\
\label{eq:LCti} LCt_{ij} =ln (cos(\alpha_{ij}))
\end{gather}

Since equation~\eqref{eq:nueve} can yield non-positive values, an additional algebraic transformation is performed prior to applying the logarithmic function. In this context, equations (18) and (19) express the variable $\theta_{ij}$ as a parametric function of the variable $0 \leq \kappa_{\theta_{i,j}} \leq 1$. For this, in equations (18) and (19) the variable $\theta_{ij}$ is represented by means of a parametric function that depends on the variable $0\leq \kappa_{\theta_{i,j}}\leq 1$, which allows $\theta_{ij}= -\widetilde{\theta}$, when $\kappa_{\theta_{ij}} = 0$ and $\theta_{ij}= \widetilde{\theta}$ when $\kappa_{\theta_{ij}} = 1$. If $\overline{\theta}= \frac{\pi}{18}$ and $\widetilde{\theta} =\frac{\pi}{6}$ it is guaranteed that $\frac{1}{3}\leq \kappa_{\theta_{i,j}}\leq \frac{2}{3}$, the region in which the logarithm is well-defined and also $\theta_{i,j}$ assumes rational values within the power flow analysis. This constraint is critical, as the angular aperture must not exceed $10^\circ$ (or $\frac{\pi}{18}$) to maintain angular stability~\cite{kundur1994}.
\begin{gather}
\label{eq:thetaij} \theta_{ij} = 2\cdot \widetilde{\theta} \cdot \kappa_{\theta_{ij}} - \widetilde{\theta} \\
-\widetilde{\theta} \leq- \overline{\theta} \leq \theta_{ij} \leq \overline{\theta} \leq \widetilde{\theta} 
\end{gather}

Considering that for the range of values of $\theta_{ij}$ it is held that $sin(\theta_{ij} )\simeq  \theta_{i,j}$, then equation \eqref{eq:nueve} can be approximated as follows:
\begin{gather}
    \label{eq:gammaij}\gamma_{ij}= 2\cdot \widetilde{\theta} \cdot \chi_{i,j}-\widetilde{\theta}\cdot w_{ij}\\
    \label{eq:chiij} \chi_{i,j} = w_{ij} \cdot \kappa_{\theta_{i,j}}\\
    \frac{1}{3} \cdot \underline{V}^{2} \leq \chi_{ij} \leq \frac{2}{3}\cdot \overline{V}^{2}
\end{gather}

Note that $\chi_{i,j}$ in (\ref{eq:gammaij}) is strictly positive and non-zero, ensuring that its logarithm is well-defined.

Hence, following the methodology outlined in equations \eqref{eq:seis} and \eqref{eq:ocho}, the nonlinearity present in equation \eqref{eq:chiij} can be address using the logarithmic transformation previously described. This approach yields:
\begin{gather}
   \label{eq:Lchi} L_{\chi_{i,j}} = L_{w_{i,j}} + L_{\kappa_{i,j}}\\
    \label{eq:Lchi1} L_{\chi_{i,j}} = ln(\chi_{i,j}) \\
    \label{eq:Lchi2} L_{\kappa_{i,j}} = ln(\kappa_{i,j}) 
\end{gather}

\begin{figure}[htbp]
\centerline{\includegraphics[width=9cm]{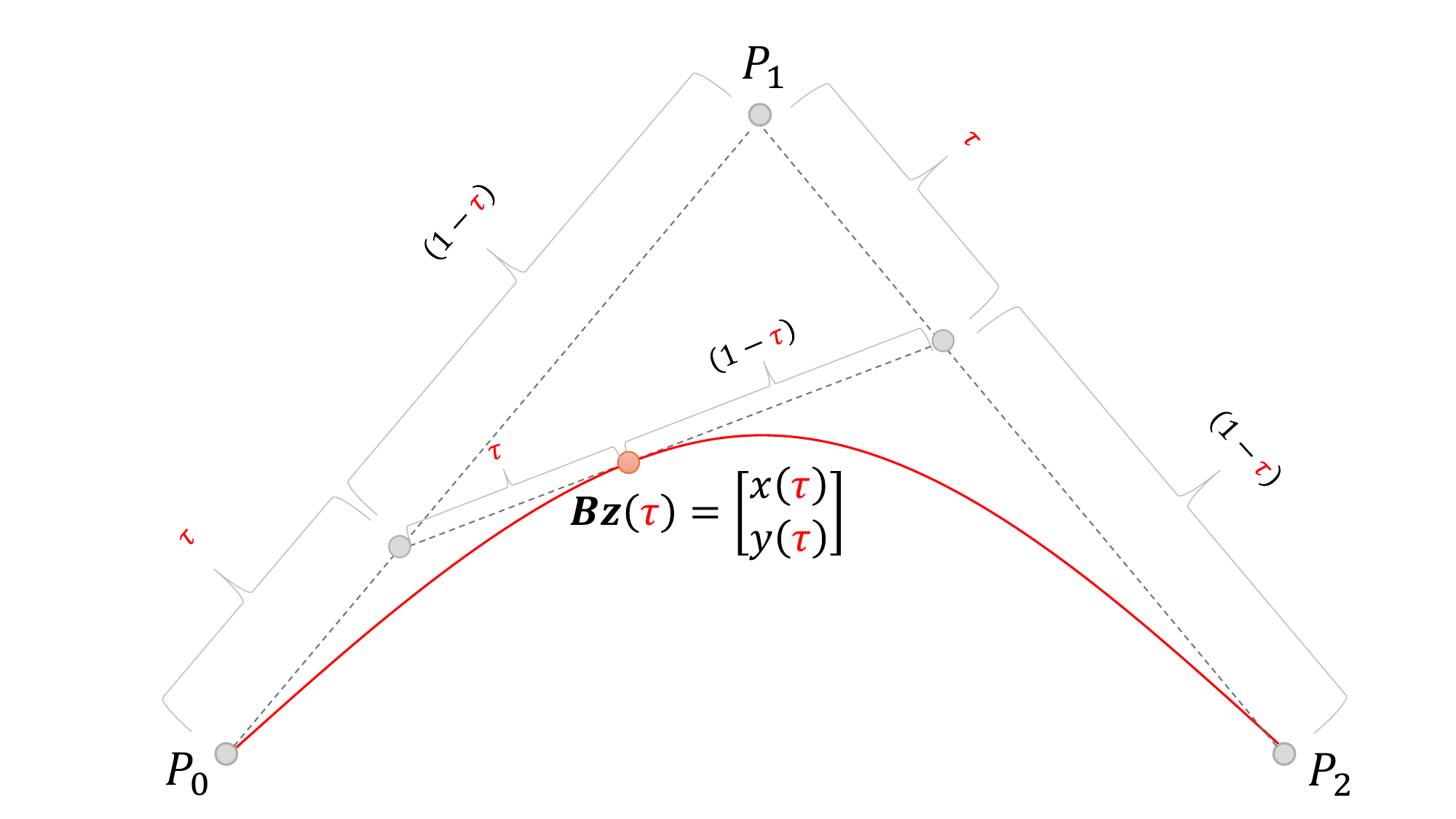}}
\caption{The second degree Bézier curves.}
\label{fig1}
\end{figure}

The equations \eqref{eq:Lphi}-\eqref{eq:Lchi2} are equivalent to the equations \eqref{eq:seis}-\eqref{eq:nueve}, with non-linearities deriving from the logarithmic functions presented in \eqref{eq:Lvi}-\eqref{eq:LCti} and \eqref{eq:Lchi1}-\eqref{eq:Lchi2}. To address these non-linearities, the logarithms are approximated by second-degree Bézier curves. These curves are generated through convex linear combinations of \(n\) control points \(P_0, P_1, \ldots, P_n\), which are defined as two-dimensional vector constants. By manipulating these control points, we can effectively model and smooth curves with complex geometries, enabling precise adjustments to achieve the desired curvature (see Fig. \ref{fig1}).

Therefore, the coordinate pairs \(\begin{bmatrix} x \\ f(x) \end{bmatrix}\) are defined in relation to the control points \(P_0 = \begin{bmatrix} x_0 \\ y_0 \end{bmatrix}\), \(P_1 = \begin{bmatrix} x_1 \\ y_1 \end{bmatrix}\), and \(P_2 = \begin{bmatrix} x_2 \\ y_2 \end{bmatrix}\). The respective parametric variable \(\tau\) varies in the interval \(0 \leq \tau \leq 1\) and is utilized to express the curve via the following parametric equation: 
\begin{gather}
\begin{split}
    \begin{bmatrix}
x \\ f(x)\end{bmatrix} \approx \textbf{Bz}(\tau) &= [P_0 \cdot (1-\tau)^{2}+2\cdot P_1 \cdot \tau\cdot (1-\tau) + \\
P_2 \cdot  \tau^{2}] &=\begin{bmatrix}
a_0+a_1 \tau +a_2 \tau^{2} \\ b_0+b_1 \tau +b_2 \tau^{2}\end{bmatrix}     
\end{split}\\
a_0 = x_0 \nonumber \\
a_1 = 2\cdot x_1 -2 \cdot x_0 \nonumber\\
a_2 = x_0 - 2\cdot x_1 -x_2 \nonumber \\
\label{eq:approx} b_0=y_0\\
b_1= 2\cdot y_1 -2 \cdot y_0 \nonumber\\
b_2= y_0- 2\cdot y_1 - y_2 \nonumber
\end{gather}

As represented in equations \eqref{eq:Lvi}-\eqref{eq:LCti} and \eqref{eq:Lchi1}-\eqref{eq:Lchi2}. The non-linear functions of interest are \( f(x) = \ln(x) \) and \( f(x) = \ln(\cos(x)) \). When these functions are approximated using second-degree Bézier curves over the relevant ranges of the variables, they produce a highly accurate fit.

By substituting equations \eqref{eq:Lvi}-\eqref{eq:LCti} and \eqref{eq:Lchi1}-\eqref{eq:Lchi2} with  their corresponding second degree Bézier approximations \eqref{eq:approx}, and executing the variable transformation, $ \xi_{(*)}=\tau^{2}_{(*)},\forall (*) \in \{ v_{i},\varphi_{i},w_{ij},\alpha_{ij},\theta_{ij},\chi_{ij},\kappa_\theta{ij} \}$, we can effectively express the solution space of the original OPF model as a highly accurate approximation through equations (2, 3, 5), (10-12), (18-20), (22, 23), and the associated Bézier representations of equations (13)-(17) and (24)-(25). This transformation allows for a more tractable analysis and solution of the OPF problem.

The compact formulation of the \textit{BF-ACOPF} model is presented in equations (28)–(33). Equation (28) corresponds to the original objective function of the ACOPF model, where $X_s$ represents all decision variables. Equation~(29) aggregates all linear constraints, including active and reactive power balances, logarithmic relations, and Bézier approximations. Equation~(30) isolates the only non-convex term—a quadratic Bézier expression—whose convexification is addressed using the BF. Finally, equations~(31)–(33) define variable bounds and restrict the Bézier parameters to the unit interval, ensuring a valid convex domain.

\begin{align}
\text{FO}_M &= 
\underbrace{\sum_s C_s \cdot X_s}_{\text{Original cost function}} 
\end{align}

Subject to:
\begin{align}
&\sum_s a_{s,h} \cdot X_s + \sum_m b_{m,h} \cdot \xi_{m(*)} + \sum_m d_{m,h} \cdot \tau_{m(*)} = b_h \\
&\tau_{m(*)}^{2} = \xi_{m(*)} \\
&\underline{X}_s \leq X_s \leq \overline{X}_s \\
&0 \leq \tau_{m(*)} \leq 1 \\
&0 \leq \xi_{m(*)} \leq 1
\end{align}

\begin{figure}[htbp]
\centerline{\includegraphics[width=9cm]{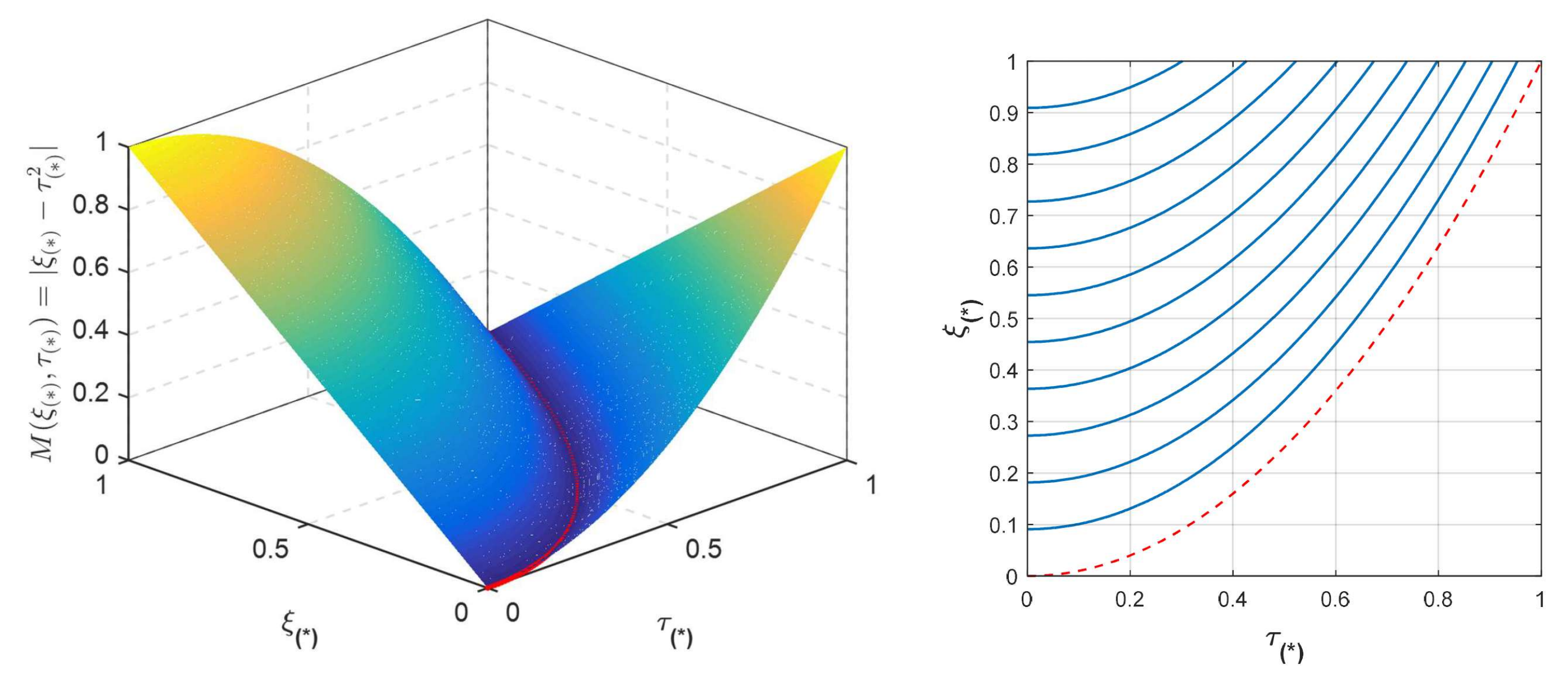}}
\caption{Butterfly-shaped function (BF).}
\label{fig2}
\end{figure}

Equation (30) is defined for the set $(*)={v_i,\varphi_i,\omega_{ij},\alpha_{ij},\theta_{ij},\chi_{ij},\kappa_{(\theta_{ij} )} }$. To convexify this equation, this work proposes reformulating it as a scalar function over \(\mathbb{R}^2\), as in (34). This function exhibits a butterfly-shaped form, as depicted in Fig. \ref{fig2}. Hence, its appellation in this work: \textit{Butterfly-shaped Function (BF)}.

\begin{equation}
    M(\xi_{(*)},\tau_{(*)})= \left| \xi_{(*)}-\tau^2_{(*)}\right|
\end{equation}

Observe in Fig. 2 (on the right) that the region in the space defined by $\xi_{(*)}$ and $\tau_{(*)}$, where equation (35) holds, is convex. Additionally, within this solution space, the function $M(\xi_{(*)},\tau_{(*)})$ depicted on the left exhibits concavity.
\begin{equation}
    \tau^2_{(*)}\leqslant \xi_{(*)}
\end{equation}

Therefore, the BF replaces equation (30) by (35). Furthermore, the term $\text{BM}\cdot(\xi_{(*)}-\tau^2_{(*)})$, is added to objective function (1), which corresponds to the BF. This results in equation (36).

\begin{gather}
\begin{split}
    FO_{M}&=\underbrace{FO}_{Equation 1}+\\
    &\underset{\forall (*)\in \{ v_{i},\varphi_{i},w_{ij},\alpha_{ij},\theta_{ij},\chi_{ij},\kappa_\theta{ij} \}}{\sum}\underbrace{\sum_{m} \text{BM}\cdot(\xi_{(*)}-\tau^2_{(*)})}_{\xi_{m(*)}=\tau_{m(*)}^{2}}
\end{split}
\end{gather}

The proposed BF-ACOPF model ensures a convex search space by design: all constraints are either linear or involve a single convex quadratic inequality. The objective function integrates a linear cost component with a concave penalization term—the BF—which is critical for ensuring both convergence and accuracy.
Despite its concave nature, the BF satisfies two crucial properties: (1) its infimum within the feasible region is zero, and (2) it is devoid of stationary points, as its gradient remains non-zero throughout. This guarantees that any optimization trajectory monotonically approaches the manifold where the function attains zero. This manifold corresponds to the pointwise fulfillment of the Bézier-based logarithmic approximations, effectively reinstating the original multiplicative nonlinearity with high accuracy, as illustrated by the red line in Fig 2.

\section{Results and discussion}
 
To validate the performance of the proposed BF-ACOPF model, tests were conducted on three systems of increasing size and complexity. Each test system highlights specific aspects of the BF-ACOPF model, such as fidelity to the exact AC model, solution quality, and computational performance. 
All optimization models were implemented in GAMS and executed on a laptop with a 13th Gen Intel\textsuperscript{\textregistered} Core\textsuperscript{TM} i5-1340P CPU (1.90~GHz) and 16~GB RAM. The B\'ezier curve parameters were tuned using a particle swarm optimizer (PSO), ensuring a maximum approximation error of $10^{-5}$. The test systems analyzed are described as follows: 

\textit{11-bus system:} A didactic system designed to analyze in detail how technical constraints impact the optimal dispatch. 

\textit{IEEE 30-bus system:} A standard test system used to directly compare the proposed BF-ACOPF model against the exact ACOPF under realistic operating conditions. It aims to evaluate the accuracy of the proposed model’s results and to show that by guaranteeing convergence to the global optimum, the BF-ACOPF model can find lower-cost solutions than the exact AC model, which is susceptible to getting trapped in local optima by conventional solvers.

\textit{IEEE 118-bus system:} A large-scale system used to validate the scalability of the BF-ACOPF model and its ability to maintain accuracy in a large network. It is verified that the solution found by the proposed model exhibits negligible error when compared with a full AC power flow simulation on the same dispatch.

\begin{figure}[htbp]
\centerline{\includegraphics[width=9.5cm]{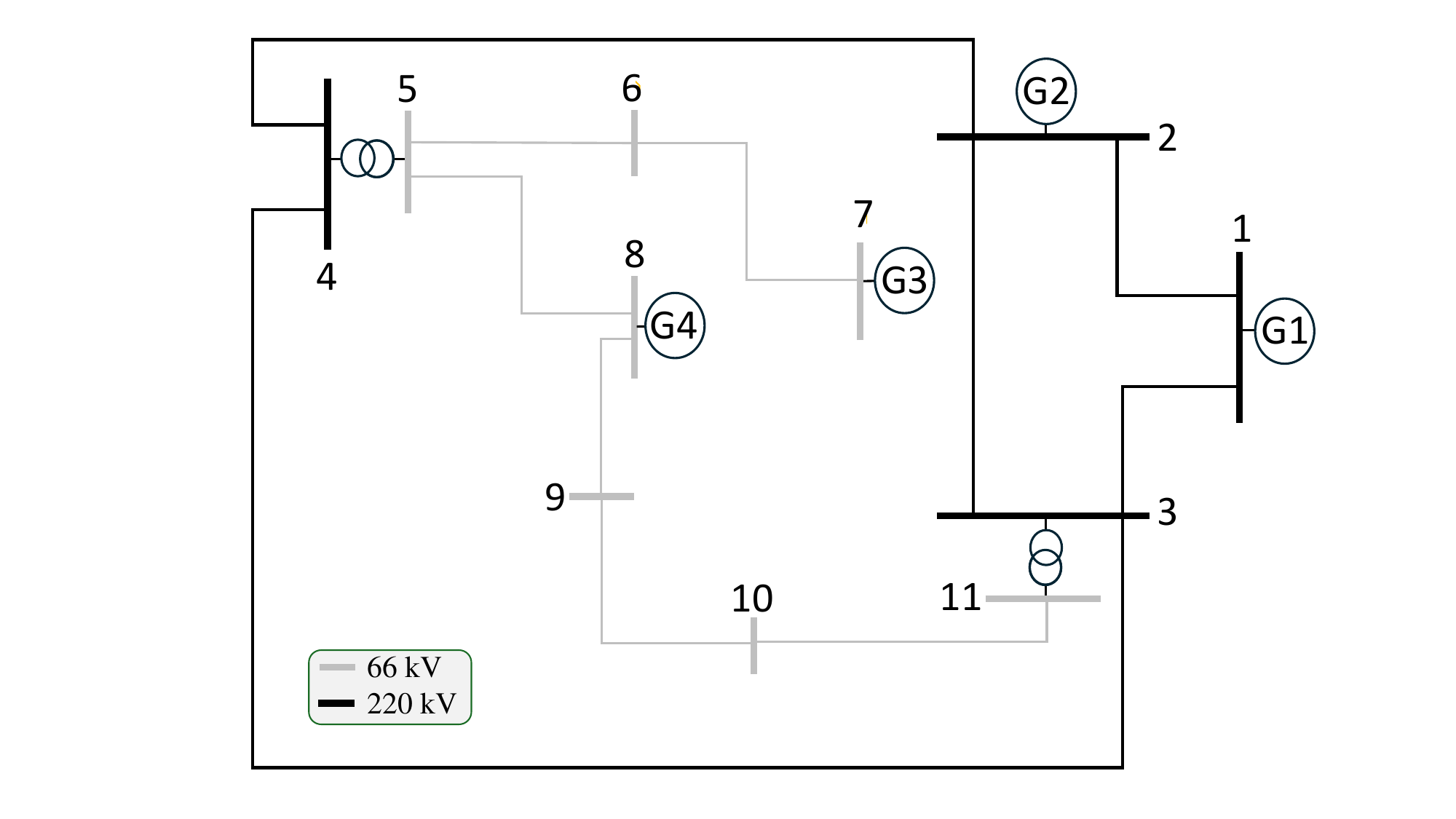}}
\caption{Single-line diagram of the 11-bus test system.}
\label{fig6}
\end{figure}

\subsection{11-Bus Test System}

The 11-bus test system (Fig.~\ref{fig6}) was inspired by the Colombian Caribbean network and carefully designed to exhibit voltage steady state issues that must be addressed through optimal dispatch. The system explicitly tests the proposed model’s ability to identify voltage constraints and to deliver a cost-minimizing dispatch that is responsive to voltage challenges. 

The network incorporates two distinct voltage levels (66~kV and 220~kV), intentionally included to examine the sensitivity of the model across varied voltage levels. It features a radial configuration at 66~kV, where the network’s topology increases its vulnerability to voltage drops at the remote ends. Furthermore, the system includes meshed connections between the two voltage levels, adding complexity with respect to power flows. In particular, congestion in the 66~kV network limits power transfer through the 230~kV level, introducing additional operational constraints that must be considered in the optimal power flow solution. Moreover, all generators in the system are characterized by detailed P--Q capability curves, fully integrated into the BF-ACOPF formulation. 

The system includes four generators: two low-cost units (G1, G2) and two high-cost units (G3, G4). To evaluate the performance of the proposed model, it is tested over four cases:

\textbf{Case 1: Single-bus (SB) dispatch.} This case ignores network flow limits and nodal voltage constraints, dispatching generation at cheapest cost.

\textbf{Case 2: DC-OPF dispatch.} A DC power flow approximation is used, accounting for line flow limits (congestion) on active power while neglecting reactive power and voltage magnitude constraints.

\textbf{Case 3: BF-ACOPF.} This convex OPF model considers all network constraints (voltage, line flows) and generator P--Q capability curves (reactive limits), however it does not include binary on/off variables (all generators are assumed available).

\textbf{Case 4: BF-ACOPF with unit commitment.} This extends Case~3 by including binary variables for generator on/off status, enforcing each generator’s minimum generation limit when committed.

Each case study isolates the impact of various operational constraints to evaluate the model's ability to identify feasible and optimal solutions. Following the dispatch for each scenario, a post-optimization AC power flow simulation is conducted using the exact AC equations to verify the physical feasibility of the solution, specifically with respect to voltage and flow limits. Table~\ref{tab:11bus-results} provides a summary of the optimal dispatch results for all cases analyzed.

\begin{table*}[ht]
\centering
\caption{Optimal dispatch results for the 11-bus system under different modeling assumptions.}
\label{tab:11bus-results}
\begin{tabular}{lcccc}
\hline
 & Case 1 (SB) & Case 2 (DC-OPF) & Case 3 (BF-ACOPF) & Case 4 (BF-ACOPF + UC) \\
\hline
Total cost (k\$)                & 68.5   & 331.5   & 404.6   & 441.5    \\
G1 (0.1~\$/kWh)                 & 635.0~MW           & 0.0~MW            & 257.4~MW; $-343.2$~MVAr & 617.9~MW; 164.9~MVAr \\
G2 (0.2~\$/kWh)                 & 25.0~MW            & 653.3~MW          & 391.3~MW; 521.7~MVAr   & 28.6~MW; 38.1~MVAr   \\
G3 (4~\$/kWh)                   & 0.0~MW             & 0.0~MW            & 7.6~MW; 10.2~MVAr      & 11.0~MW$^*$; 10.1~MVAr \\
G4 (30~\$/kWh)                  & 0.0~MW             & 6.7~MW            & 9.0~MW; 12.0~MVAr      & 11.0~MW$^*$; 14.7~MVAr \\
Total generation (MW)          & 660.0             & 660.0           & 665.3              & 668.5             \\
Feasible under AC simulation   & No                & No              & Yes                & Yes               \\
\hline
\multicolumn{5}{l}{$^*$Minimum technical generation limit.}
\end{tabular}
\end{table*}

The Case~1 dispatch reflects the optimal cost scenario for meeting demand under the assumption of no physical network constraints. This setup prioritizes economic efficiency, leading to the full output of the least expensive generator G1 at 635 MW, with the remaining demand met by G2 at 25 MW, yielding the lowest possible cost. However, this solution disregards essential network and voltage constraints, rendering it physically infeasible for implementation in real-world scenarios.

In Case 2, the DC-OPF model results in a more expensive dispatch strategy that takes network constraints into account. Notably, congestion occurs on the line connecting nodes 10--11, with the flow reaching the operational limit of 23~MW.

While the Case~2 solution is more realistic than Case~1 given that it reflects the impact of active power flow limits, remains impractical because it disregards voltage magnitude and reactive power demands. 
A subsequent AC power flow analysis reveals that the dispatch configuration from Case~2 cannot meet the voltage constraints without either exceeding the reactive power limits of the generators or requiring load shedding. Consequently, it is deemed infeasible when evaluated under the precise AC model.

In Case~3, the implementation of the proposed BF-ACOPF model leads to a significantly different generation dispatch compared to Case~2. The BF-ACOPF solution accounts not only for reactive power requirements for admissible voltage levels but also for power losses and constraints imposed by generator capability curves. Consequently, the proposed model identifies the need to activate the higher-cost generator G3 (7.6~MW, supplying 10.2~MVAr) to provide voltage support at bus~7, located at the far end of the radial 66~kV subsystem.

In Case 4, the BF-ACOPF approach incorporates unit commitment binary variables, yielding a fully operational dispatch that satisfies all network and generator constraints. This model effectively commits the necessary generators while enforcing their minimum output levels, resulting in a cost-optimal solution that meets all steady-state power system requirements. The dispatch outcomes across Cases 1 to 4 exhibit significant variation, highlighting the sensitivity of the OPF problem to the integration of network constraints. Notably, Case 4 (BF-ACOPF with unit commitment) achieves a feasible, optimal dispatch that satisfies the AC power flow equations and all equipment limits, compared to the other cases. This solution properly manages reactive power injections to maintain voltage stability, accounts for network losses, and considers each unit's generator capability curves, along with its operational maximum and minimum output thresholds.

In summary, the results from the 11-bus test system illustrate the critical influence of network flow limitations, voltage regulation constraints, power losses, and the co-optimization of active and reactive power on optimal economic dispatch. Within this framework, the BF-ACOPF model emerges as a robust analytical tool for tackling these complexities, yielding globally optimal solutions that closely align with the real physical operation of the power system.

\subsection{IEEE 30-Bus System}
The IEEE 30-bus system contains 37 transmission lines and 4 transformers, with 6 generators located at buses 1, 2, 13, 23, 24 (slack), and 27. Additionally, a synchronous condenser (fixed $P=0$) of $\pm 50$~MVAr capacity is connected at bus~30 to provide reactive support.

When using simplified models such as DC-OPF or the BF-ACOPF, it is imperative to perform a post-validation step to ensure the solution is physically feasible. This AC validation consists of fixing the state variables (specifically, the dispatch generator set-points: all generator active powers, all PV-bus voltages, and the slack bus voltage magnitude and angle) as given by the OPF solution, and then running a power flow simulation using the full AC equations. This determines whether the OPF solution can be adjusted to a feasible AC operating point and quantifies any necessary corrections (such as slack generation adjustments or load shedding). 

Three variants of the OPF were solved in the 30-bus system, incorporating binary commitment decisions for generators: (i) DC-OPF, (ii) BF-ACOPF, and (iii) the exact AC-OPF (non-convex). Table~\ref{tab:30bus-compare} presents the dispatch results for the DC-OPF and BF-ACOPF models along with their AC post-validation outcomes, and for the exact ACOPF solution. In table ~\ref{tab:30bus-compare}, “AC Sim.” denotes the AC power flow simulation results when the corresponding OPF dispatch is implemented. For each generator (G1--G6) and the synchronous condenser (SC), the dispatched active power ($P_g$) and reactive power ($Q_g$) are shown. The row “DNS” indicates the amount of demand not served (load shed), if any, needed to achieve an AC feasible solution. “Total” gives the total generation dispatched, and “Losses” is the total system losses. The generation cost for each case is also reported.

\begin{table*}[ht]
\centering
\caption{Dispatch comparison of DC-OPF, BF-ACOPF, and exact AC-OPF for the IEEE 30-bus system (all with generator on/off optimization).}
\label{tab:30bus-compare}
\begin{tabular}{lcccccccccc}
\hline
 & \multicolumn{2}{c}{DC-OPF} & \multicolumn{2}{c}{AC Sim. of DC-OPF} & \multicolumn{2}{c}{BF-ACOPF} & \multicolumn{2}{c}{AC Sim. of BF-ACOPF} & \multicolumn{2}{c}{Exact AC-OPF} \\
 & $P$ (MW) & $Q$ (MVAr) & $P$ (MW) & $Q$ (MVAr) & $P$ (MW) & $Q$ (MVAr) & $P$ (MW) & $Q$ (MVAr) & $P$ (MW) & $Q$ (MVAr) \\
\hline
G1 (0.9~\$/kWh)        & 147.00 & --    & 147.00 & 7.36   & 147.00 & $-7.36$  & 147.00 & $-7.15$  & 144.68 & 30.46  \\
G2 (20~\$/kWh)       & 0.00   & --    & 0.00   & 0.00   & 10.97  & 46.06   & 10.97  & 47.36   & 0.00   & 0.00   \\
G3 (30~\$/kWh)& 67.00  & --    & 52.48  & 27.43  & 65.00  & 29.50   & 65.47  & 29.81   & 77.89  & 28.49  \\
G4 (5~\$/kWh)       & 0.00   & --    & 0.00   & 0.00   & 0.00   & 0.00    & 0.00   & 0.00    & 0.00   & 0.00   \\
G5 (15~\$/kWh)       & 0.00   & --    & 0.00   & 0.00   & 0.00   & 0.00    & 0.00   & 0.00    & 0.00   & 0.00   \\
G6 (1~\$/kWh)       & 65.00  & --    & 65.00  & 33.64  & 65.00  & 25.32   & 65.00  & 25.79   & 65.00  & 33.64  \\
SC (bus 30) & 0.00 & --    & 0.00   & 20.16  & 0.00   & 16.11   & 0.00   & 15.18   & 0.00   & 17.43  \\
DNS       & 0.00   & --    & 19.56  & --     & 0.00   & --      & 0.00   & --      & 0.00   & --     \\
Total     & 279.00 & --    & 264.48 & 88.58  & 287.98 & 93.52   & 288.44 & 95.82   & 287.57 & 92.59  \\
Losses    & 0.00   & --    & 5.05   & $-11.10$ & 8.98  & $-13.68$ & 9.44  & $-11.38$ & 8.57  & $-14.61$ \\
Gen. cost (k\$) & 2207.30 & --  & 3728.28 & --   & 2366.86 & --    & 2380.83 & --    & 2531.89 & --    \\
\hline
\multicolumn{11}{l}{Note: ``--'' indicates a value not applicable to that case (e.g. reactive power in DC-OPF, or Q for shed load).}
\end{tabular}
\end{table*}

\subsubsection{Performance of the DC-OPF solution}
The DC-OPF model produces a dispatch that maximally leverages the most cost-effective generation units, specifically G1 and G6, operating them up to their capacity limits (refer to Table~\ref{tab:30bus-compare} for details). However, when this dispatch is subjected to AC validation, it is found to be infeasible without incurring unserved demand. The AC simulation of the DC-OPF dispatch could not find a solution that adheres to the required voltage and reactive power constraints without shedding 19.56 MW of load, as noted in the DNS row. As a result, the adjusted operating condition incurs significant penalties, reflected in a substantial increase in generation costs to k\$~3728.28 post-AC adjustment, compared to the original k\$~2207.30.

\subsubsection{Performance of the BF-ACOPF model}
The BF-ACOPF model identifies an optimal dispatch that closely approximates to a feasible solution within the AC framework, requiring slight modifications during post-validation.

As shown in Table~\ref{tab:30bus-compare}, the BF-ACOPF dispatch makes minor adjustments to the output of the slack generator (G3’s $P$ adjusted from 65.00 to 65.47~MW) during the AC power flow simulation, with no need for load shedding. The adjusted AC solution closely mirrors the original BF-ACOPF dispatch, suggesting a high degree of alignment with the actual AC power flow equations. Notably, the BF-ACOPF solution does not involve any DNS (unserved demand) and adheres to all operational constraints when evaluated against the precise AC model.

\begin{figure}[ht]
    \centerline{\includegraphics[width=8cm]{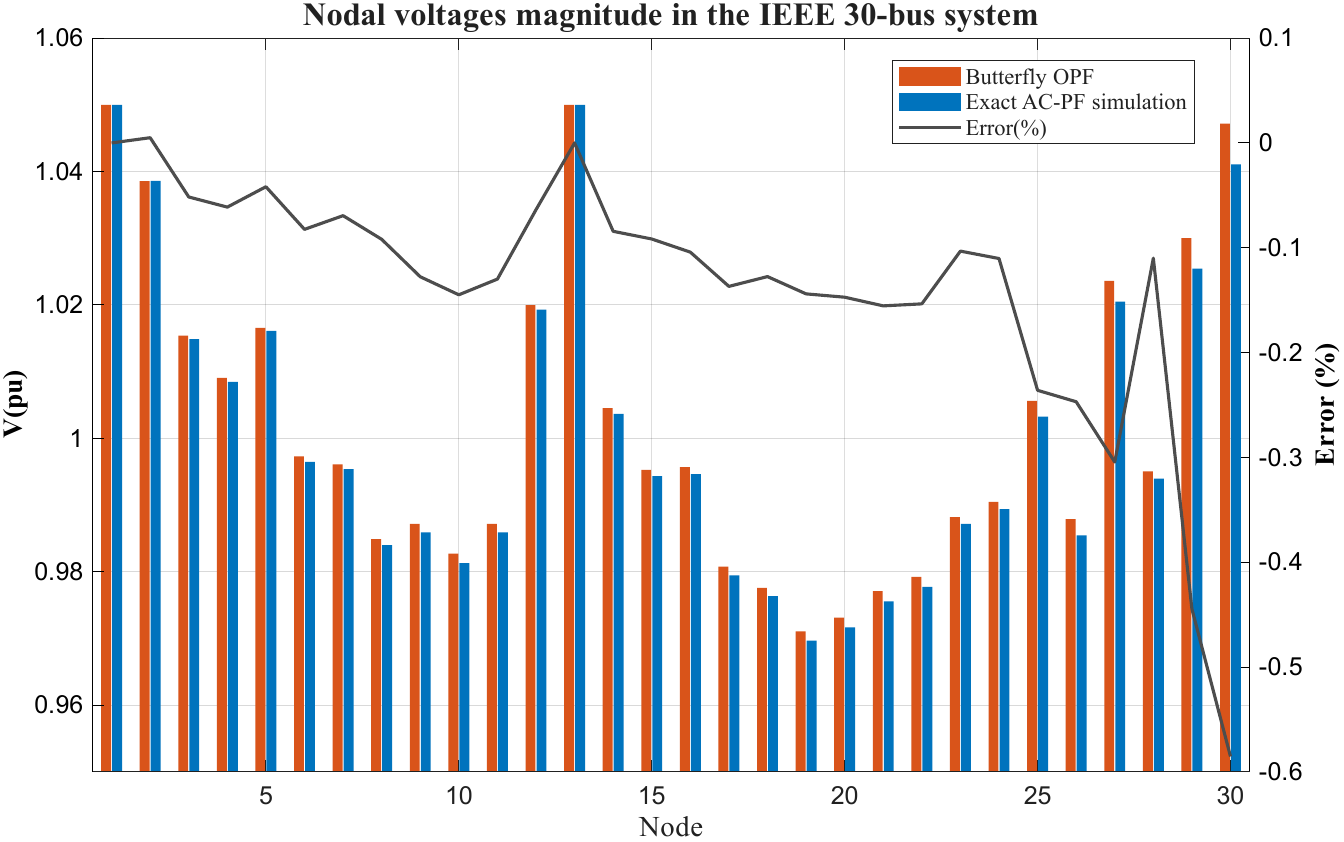}}
    \centerline{\includegraphics[width=8cm]{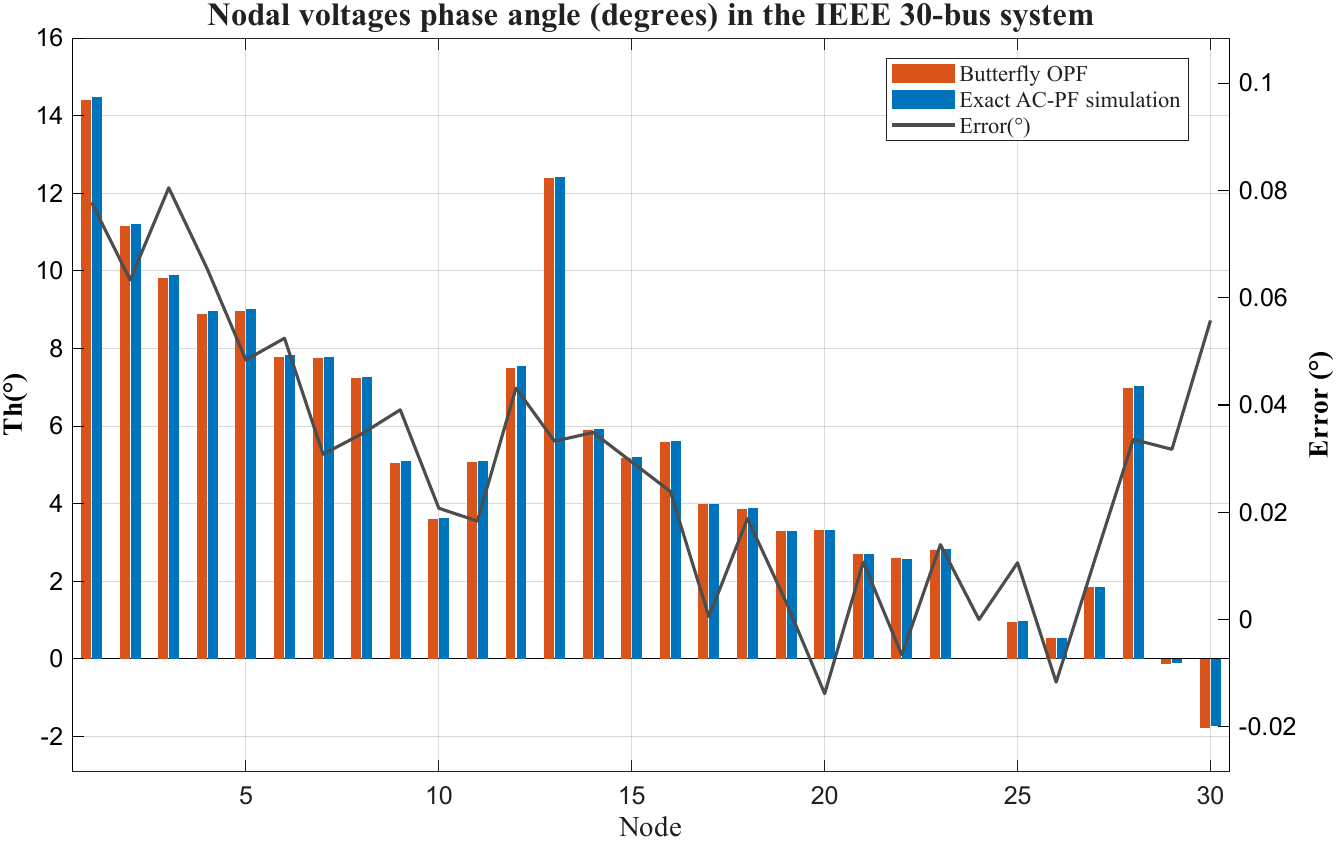}}
    \caption{Nodal voltages in the IEEE 30-bus system: (a) voltage magnitude (pu); (b) phase angle (degrees). Orange bars: BF-ACOPF OPF solution; blue bars: exact AC power flow simulation.}
    \label{fig:30bus-voltage}
\end{figure}

\begin{figure}[ht]
    \centerline{\includegraphics[width=8cm]{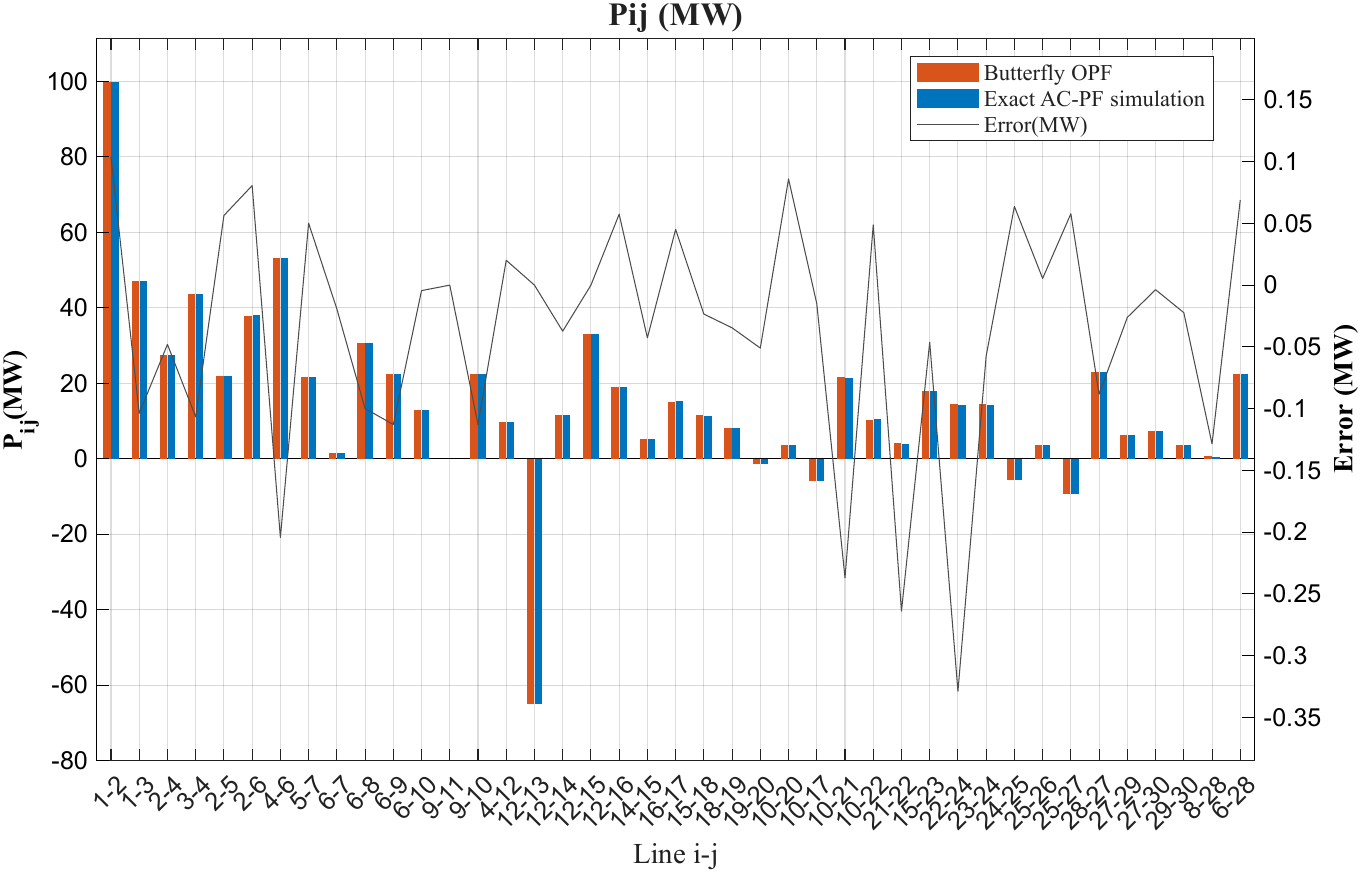}}
    \centerline{\includegraphics[width=8cm]{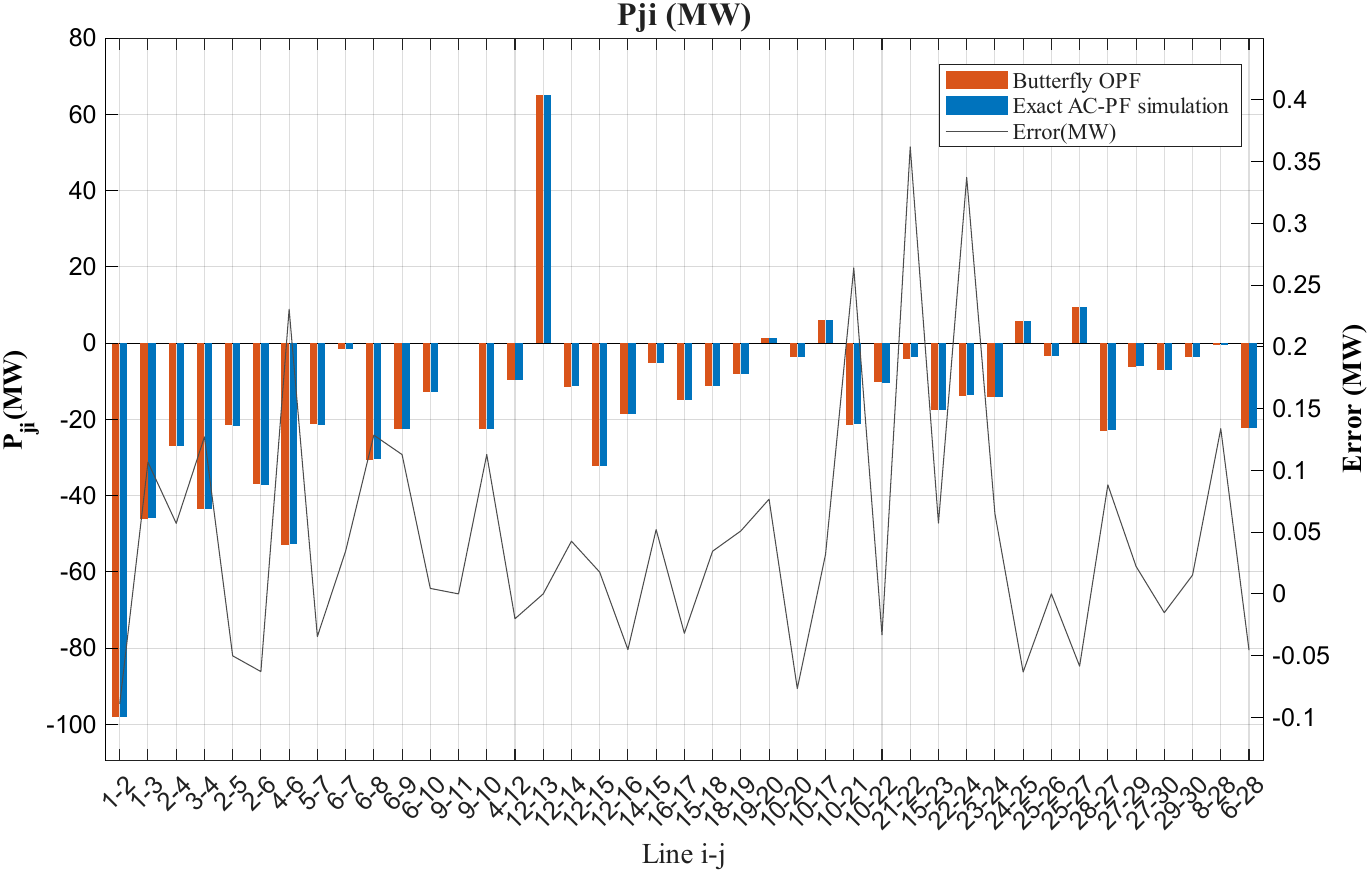}}
    \caption{Branch active power flows in the IEEE 30-bus system: (a) $P_{ij}$ (MW); (b) $P_{ji}$ (MW). Orange: BF-ACOPF OPF; Blue: exact AC.}
    \label{fig:30bus-Pflow}
\end{figure}

\begin{figure}[ht]
    \centerline{\includegraphics[width=8cm]{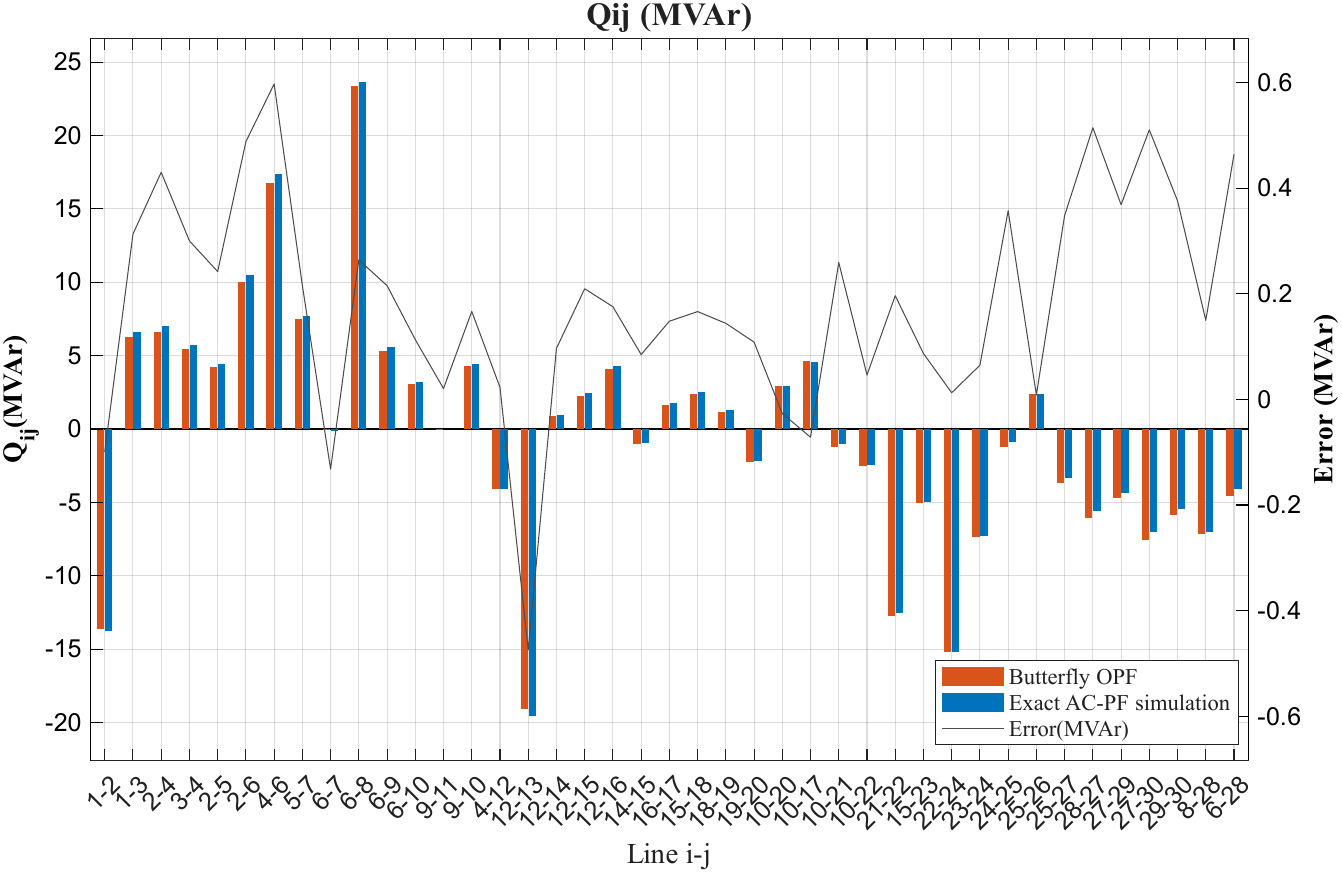}}
    \centerline{\includegraphics[width=8cm]{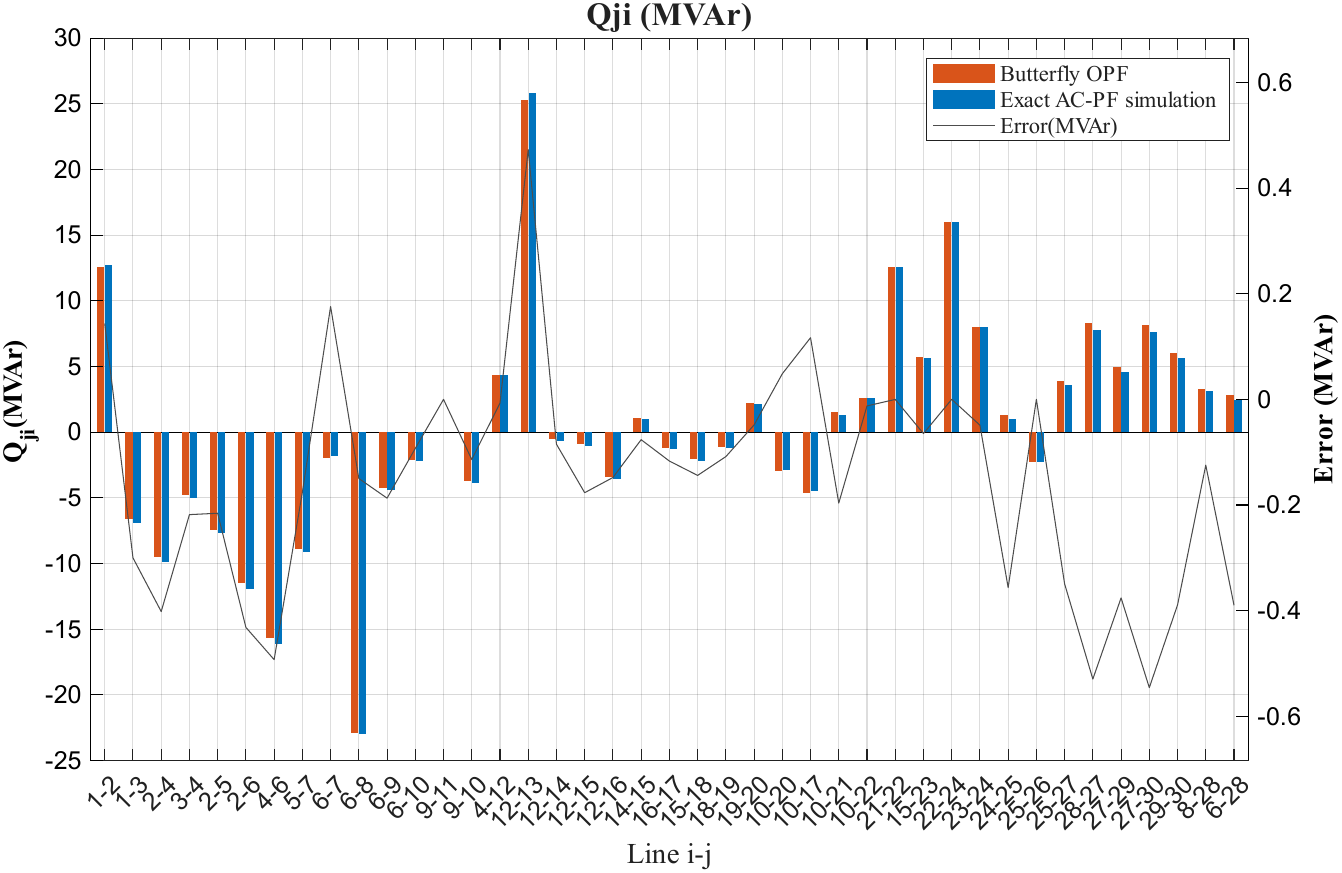}}
    \caption{Branch reactive power flows in the IEEE 30-bus system: (a) $Q_{ij}$ (MVAr); (b) $Q_{ji}$ (MVAr). Orange: BF-ACOPF OPF; Blue: exact AC.}
    \label{fig:30bus-Qflow}
\end{figure}

The fidelity of the BF-ACOPF model in relation to exact AC physics is quantitatively demonstrated through minimal errors in state variables and power flows. Specifically, the average error in nodal voltage magnitudes between the BF-ACOPF solution and the exact AC validation is 0.14\%, while the average phase angle error is 0.03 degrees. As illustrated in Figure~\ref{fig:30bus-voltage}, the nodal voltage profiles derived from the proposed model are indistinguishable from those obtained through exact AC power flow analyses, reflecting highly accurate results for both voltage magnitudes and angles.
Furthermore, the power flows derived from the BF-ACOPF model exhibit excellent alignment with the exact AC flows: the mean active power flow error was 0.07 MW in the $i \rightarrow j$ direction and 0.08 MW in the $j \rightarrow i$ direction. For reactive power flows, the mean errors were 0.22 MVAr ($i \rightarrow j$) and 0.20 MVAr ($j \rightarrow i$). Figures~\ref{fig:30bus-Pflow} and \ref{fig:30bus-Qflow} further validate that both active and reactive branch flow values computed from the proposed model align with high accuracy to those generated by the exact AC power flow model.

The BF-ACOPF model's solution demonstrated superior performance, achieving a lower objective value than the exact AC-OPF model. Specifically, the cost identified by the proposed model was \$2380.83, followed by an AC validation, while the exact AC-OPF reported a cost of \$2531.89, which fell into a local minimum. This discrepancy of approximately \$151.05 reflects a 6.34\% deviation from the global optimum cost. Notably, the non-convex nature of the exact AC formulation led the solver to a dispatch that was roughly 6.3\% more expensive than the global minimum. In contrast, the convex BF-ACOPF model successfully ensured convergence to the global optimum, effectively circumventing local optima inherent in the exact AC-OPF. This illustrates a significant advantage of the proposed approach: it not only identifies feasible solutions but also consistently delivers lower-cost outcomes.

\subsubsection{Comparison with the exact AC-OPF solution}
The findings underscore significant differences between the dispatch outcomes generated by the exact AC-OPF model and those produced by the proposed model. The differences are significant for generators G1, G2, and G3. They can be attributed to the network's operational constraints (i.e., reactive power requirements and capability curve) and to the non-convexity inherent in the exact AC formulation.

The conventional exact formulation of AC-OPF is inherently nonconvex, making it susceptible to multiple local optima. As highlighted in Table II, the exact model converges to a suboptimal solution in which Generator G1, with the lowest production cost, operates at an active power output of 144.68 MW. This underutilization occurs because a portion of its capacity (30.46 MVAr) is allocated to reactive power production to meet network constraints. In contrast, the proposed convex model effectively avoids local optima, enabling a reactive power dispatch strategy that not only identifies a feasible operating point but also maximizes active power extraction from the most cost-efficient generators.

Given that the exact AC-OPF model converged to a local optimum, it failed to sufficiently reduce the output of the most expensive generator, G3. Furthermore, it did not dispatch generators G1, G2, and G3 optimally, which could have effectively contributed to reactive power support and enabled a more economical redistribution of active power. Fundamentally, the exact AC solution did not identify the configuration that minimizes reliance on G3 by leveraging the less expensive capacity of G2 and the other generators to efficiently satisfy both active and reactive power requirements.

The proposed model, due to its convexity, finds a feasible globally optimal dispatch that efficiently allocates generation resources. In the BF-ACOPF solution, G2 operates near its minimum output to provide the required reactive power for voltage support. This strategic dispatch allows G1 to operate at levels close to its maximum active power output, effectively leveraging G1's low operational costs while adhering to its reactive capability limits. Consequently, G2's generation effectively displaces a substantial portion of G3's output, thereby reducing overall operational costs.

The results derived from the 30-bus system demonstrate that the proposed model successfully enhances resource allocation efficiency. The optimization leads to dispatch solutions that are both operationally viable and economically beneficial.

\subsection{IEEE 118-Bus System}
\begin{figure}[ht]
    \centerline{\includegraphics[width=8cm]{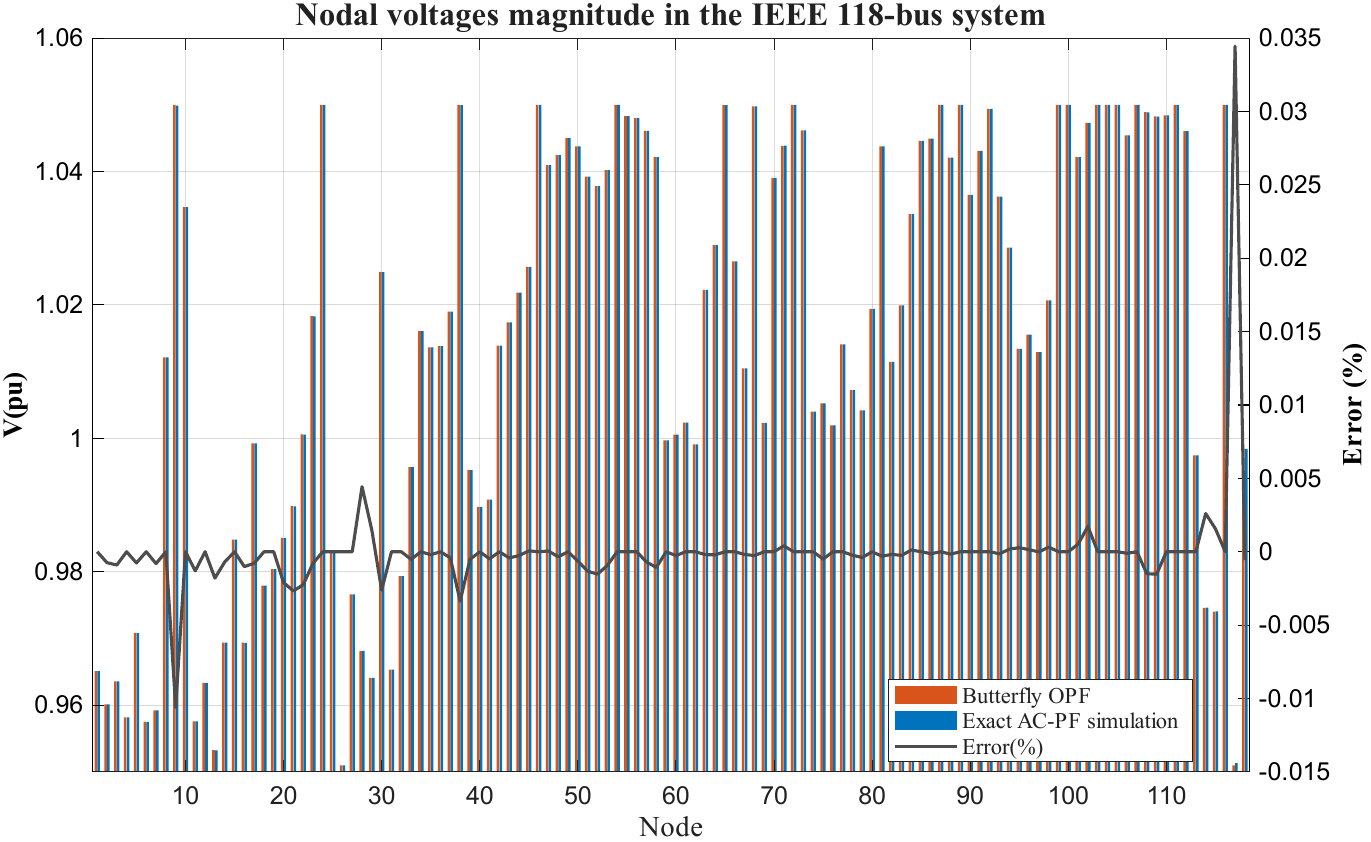}}
     \centerline{\includegraphics[width=8cm]{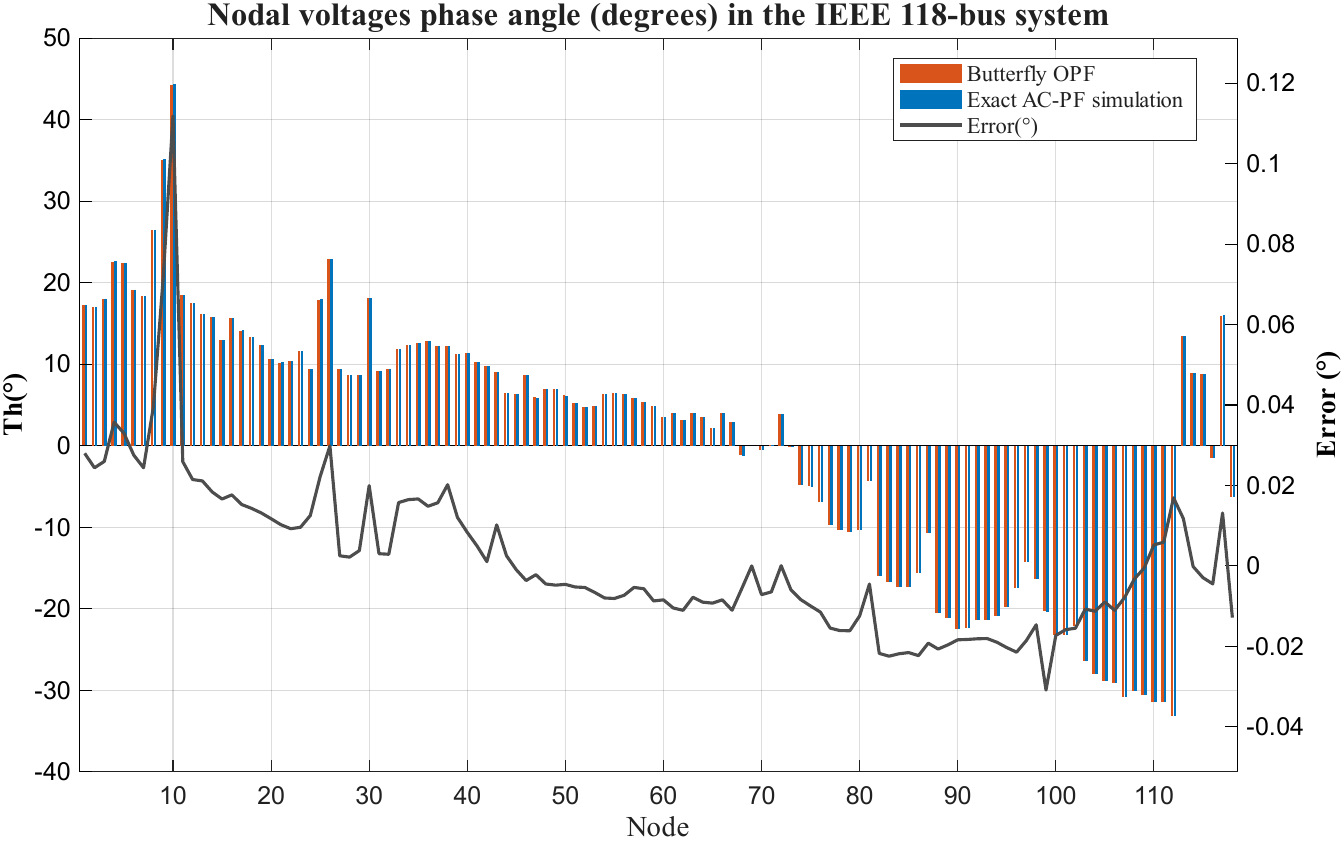}}
    \caption{OPF results for the IEEE 118-bus system, comparing the BF-ACOPF solution and exact AC power flow validation: (a) nodal voltage magnitudes (pu); (b) nodal voltage phase angles (degrees).}
    \label{fig:118bus-voltage}
\end{figure}

The IEEE 118-bus system is used to assess the scalability and accuracy of the proposed model. The OPF problem for this network was solved using discrete binary decisions for generator status within the BF-ACOPF formulation. The accuracy of the solution is evaluated by fixing the generator set-points (active powers and voltage set-points) at the BF-ACOPF optimal values and simulating the system using an exact AC load flow. Figure~\ref{fig:118bus-voltage} shows the results of the proposed model (in orange bars) and the AC load flow validation (in blue bars). The voltage magnitudes and phase angles exhibit strong similarity across all 118 buses evaluated in the tests. The maximum observed error for voltage magnitudes is a of 0.035\%, while the maximum deviation in phase angles is of $0.11^\circ$. Moreover, the average percentage errors in the state variables—specifically, the magnitudes and angles of nodal voltages—of just 0.0008\% and $0.014^\circ$, respectively, when compared with the precise AC formulation. These findings validate that the proposed model exhibits high fidelity to the exact AC equations in a large-scale system such as the IEEE 118-bus network.

The 118-bus BF-ACOPF model comprises 6,433 continuous and 54 binary variables, achieving rapid convergence in just 8.11 seconds. This outcome underscores the method's effectiveness in terms of optimality, accuracy, and computational efficiency. Such rapid performance is crucial for real-world power system operations and planning, where timely solutions are essential. In summary, the proposed model demonstrates remarkable scalability and consistently yields high-quality, feasible solutions that closely match the exact AC power flow results, while maintaining global optimality for the OPF problem.

\subsection{Key Findings}
The results from the four analyzed cases demonstrate completely different dispatches, highlighting the sensitivity of the optimal power flow problem to different operational constraints. Only Case 4 (AC-OPF with binary commitment variables) achieves an optimal and feasible dispatch that meets the steady-state power system conditions. This is because reactive power injection must be managed to maintain voltage levels, in addition to considering network losses and the influence of generator capability curves, along with their minimum and maximum generation limits.
These findings show that factors such as line loading limits, voltage constraints, network losses, and the proper dispatch of active and reactive power significantly impact economic optimization results. In this context, the BF-ACOPF Model proves to be an effective tool for addressing these challenges, providing globally optimal solutions that closely match the real operational behavior of the power system.

\section{Conclusions}
This work presents a convex reformulation of the AC optimal power flow problem by introducing auxiliary variables to isolate nonlinearities, applying logarithmic transformations to leverage product-sum properties, and employing Bézier curve approximations guided by a novel convexifying butterfly-shaped function. 

The model was numerically validated, exhibiting high fidelity to the exact AC formulation while outperforming non-convex approaches in both solution quality and computational efficiency. In simulations conducted on the IEEE 118-bus system, the proposed model achieved average errors of only 0.0008\% in voltage magnitudes and 0.014° in voltage angles, with an optimal solution computed in just 8.11 seconds. 

Furthermore, the model preserves strong sensitivity between decision variables and key electrical phenomena such as reactive power support, transmission losses, and voltage levels, ensuring physical consistency without compromising convexity or global optimality. These results position the proposed approach as a robust and practical tool for addressing the technical and operational challenges of modern power systems, especially for those weak power systems that face problems with reactive power supply and overall grid robustness.

\bibliographystyle{IEEEtran}
\bibliography{biblio.bib}

\vfill

\end{document}